# Phase variance as a seismic quality-control attribute
*Akshika Rohatgi, Andrey Bakulin, and Sergey Fomel*

**ABSTRACT**

Seismic wavefields recorded on land are strongly distorted by near-surface heterogeneity, which introduces trace-specific, frequency-dependent phase perturbations that persist even after advanced time processing. Conventional processing relies primarily on surface-consistent deconvolution, which targets long- to mid-wavelength phase variability under the highly simplified assumption of surface consistency, equalizing large-scale trends and taming variability through overdetermination. However, this approximation is inherently unable to correct localized, non-surface-consistent phase distortions, and its effectiveness further degrades when such effects dominate, as is often the case for point-receiver data.

A separate and equally important limitation is that conventional workflows provide no direct, quantitative measure of phase reliability. Phase quality is therefore assessed only indirectly, typically through amplitude behavior or visual inspection, leaving residual phase disorder largely undiagnosed.

We introduce phase variance as a seismic quality-control attribute by treating seismic phases as circular random variables and analyzing local trace ensembles using circular statistics. This data-driven measure quantifies localized phase dispersion without phase unwrapping, enabling analysis of local phase trends and fluctuations without global assumptions or wavelet models. Phase variance is computed automatically and provides frequency-by-frequency classification of the data, ranging from coherent signal behavior to fully randomized, noise-dominated phase. Synthetic tests confirm that phase variance reliably captures imposed phase perturbations and their frequency dependence.

Application of phase variance analysis to field prestack land data shows that conventional processing reduces phase variability primarily in the low-to-intermediate frequency range and struggles within the noise cone, while the highest and lowest frequencies often show little improvement in phase coherence. Phase variance operates automatically over the full prestack volume, from shallow to deep, and frequency by frequency, providing a consistent, human-independent metric for defining effective bandwidth based on phase coherence and supporting phase-sensitive workflows such as AVO, migration, and full-waveform inversion.

**INTRODUCTION**

Seismic wavefields recorded at the surface are inevitably distorted by near-surface heterogeneity and small-scale elastic contrasts. These perturbations scatter energy and introduce complex, frequency-dependent distortions of amplitude and phase (Sato, Fehler, and Maeda, 2012). In land seismic processing, such effects have traditionally been mitigated through the extensive use of source and receiver arrays during acquisition (Meunier, 2011), followed by surface-consistent processing methods. Surface-consistent deconvolution and residual statics model the data using operators assigned separately to sources and receivers, implicitly assuming that near-surface distortions repeat in a predictable manner across traces (Taner, Koehler, and Alhilali, 1974; Taner and Koehler, 1981). Randomness is accommodated only in a limited sense, typically through trace-dependent static time shifts parameterized by one operator per source and one per receiver. These methods were deliberately designed to limit the number of degrees of freedom and overdetermine the estimation problem, enabling robust first-order corrections for statics and coupling, but without being grounded in a physical model of wave propagation through heterogeneous media.

Cary and Nagarajappa (2014a, 2014b) explicitly demonstrated that surface-consistent deconvolution can leave substantial residual phase errors and proposed additional surface-consistent phase corrections coupled with residual statics estimation. However, the underlying assumption remained that phase corrections are frequency-independent and surface-consistent. As a result, these approaches remain fundamentally unable to address the trace-to-trace, frequency-dependent phase variability that is abundant in modern single-sensor data. Several extensions have sought to relax strict surface consistency by introducing non-surface-consistent, time-varying statics that allow trace-dependent perturbations (Reilly et al., 2010). While these methods come closest to acknowledging the increased complexity of near-surface effects within conventional processing, they remain fundamentally constrained. By construction, they address only travel-time corrections and do not account for localized, frequency-dependent phase perturbations arising from small-scale scattering. Consequently, the available degrees of freedom remain insufficient to represent waveform distortions that vary continuously along the trace and differ for each source–receiver pair.

In practice, wave propagation through heterogeneous near-surface media is not surface consistent. Small-scale heterogeneity induces forward scattering and interference that generate non–surface-consistent, trace-specific phase perturbations (Bakulin et al., 2022a). These perturbations are not purely kinematic and cannot be corrected by time shifts alone. Instead, they modify the phase locally in time and frequency, even after advanced statics and deconvolution have been applied. As a result, residual phase disorder persists throughout prestack gathers, revealing a fundamental gap between classical processing assumptions and the physics of wave propagation in heterogeneous media (Stork et al., 2020; Bakulin et al., 2020).

Ensemble-based processing methods implicitly acknowledge the importance of phase coherence by exploiting redundancy in multi-channel seismic data. Historically, this redundancy was realized directly in acquisition through the use of source and receiver arrays, where local stacking in the time domain improved phase coherence but simultaneously conditioned phase and amplitude, a limitation perceived as detrimental to amplitude fidelity and high-frequency preservation (Newman and Mahoney, 1973). In seismology, phase-weighted stacks were introduced to explicitly prioritize phase coherence for signal detection, reflecting the recognized outsized importance of recovering correct phase from multi-channel recordings (Schimmel and Paulssen, 1997; Schimmel and Gallart, 2007). Phase-estimation and enhancement methods based on local attributes and blind deconvolution were later proposed to stabilize phase behavior (van der Baan, 2008; van der Baan and Fomel, 2009; Fomel and van der Baan, 2014; Holt and Lubrano, 2020). More recently, with the widespread adoption of point-receiver acquisition, ensemble-based phase conditioning has been reintroduced into reflection seismic processing through seismic time-frequency masking, where phase manipulation is decoupled from amplitude effects (Bakulin et al., 2023). While these methods effectively reduce incoherent phase locally, they treat phase variability primarily as a nuisance to be suppressed and do not explicitly analyze phase as a statistical quantity or quantify the underlying phase distributions that govern coherence loss.

Despite decades of advances in noise attenuation and imaging, and long-standing recognition of the critical role of phase coherence, seismic processing still lacks a routine, quantitative, and objective measure of phase reliability. In practice, phase quality is assessed indirectly through visual inspection, stack response, or amplitude-based attributes, relying on ad-hoc heuristics and human judgment rather than a standardized metric that isolates phase behavior or captures its frequency-dependent variability. This limitation is most evident in land environments with strong near-surface heterogeneity, where pronounced phase variability persists despite careful conventional processing (Bakulin et al., 2020; Stork, 2020). The problem becomes acute in desert environments such as the Middle East, where near-surface heterogeneity is particularly strong (Bakulin et al., 2022a; 2024), and reaches its extreme in highly scattering media such as basalts and crystalline rocks, where seismic imaging is often restricted to very low frequencies (Ziolkowski et al., 2003).

Recent work has reframed near-surface scattering as a multiplicative process that randomizes seismic phase, giving rise to speckle-like behavior in seismic wavefields (Bakulin et al., 2022a; 2023; Rohatgi et al., 2025). Analogous phenomena are well documented in optics and acoustics, where volumetric scatterers produce frequency-dependent phase fluctuations and coherence loss that are best described statistically (Goodman, 2007; Abbott and Thurstone, 1979). In seismic data, forward scattering on meter-scale heterogeneity generates multiple near-ballistic arrivals whose interference leads to rapid, trace-to-trace phase variations. These phase perturbations often dominate waveform degradation and increase with frequency, driving the wavefield toward a speckle regime in which individual traces appear chaotic, yet ensemble statistics remain stable and physically meaningful.

Modern land seismic acquisition is characterized by dense spatial sampling with point receivers rather than source-receiver arrays. While this increases spatial resolution and preserves high-frequency information, it also makes small-scale near-surface heterogeneity and trace-to-trace phase variability far more apparent within localized, densely sampled ensembles. In such settings, propagation through heterogeneous near-surface media renders the observed seismic phase effectively random from trace to trace.

This study adopts an explicitly stochastic viewpoint. Although the seismic source is deterministic, the resulting wavefield observed in prestack data reflects the cumulative effects of near-surface scattering and interference. Importantly, seismic phase is inherently circular rather than linear and must be treated accordingly. We introduce an ensemble-based framework that models seismic phase as a circular random variable and applies circular statistics to quantify phase variability within prestack gathers. We define a phase-variance attribute that summarizes frequency-dependent phase disorder. By shifting from single-trace phase estimates to ensemble phase distributions, the proposed framework provides a physically grounded measure of scattering-induced phase noise and establishes a transparent link among near-surface heterogeneity, phase disorder, and seismic coherence.

**PHASE BEHAVIOR AND STATISTICS IN SEISMIC ANALYSIS**

To motivate a statistical description of seismic phase, we start with controlled examples that show how phase coherence changes within local trace ensembles. Rather than interpreting phase trace by trace, we treat the phases at each frequency as a sample drawn from an underlying distribution and track how that distribution evolves as noise increases. We therefore analyze a sequence of cases spanning the full range of signal-to-noise ratios (SNR), anchored by two end members, a pure signal reference and a pure noise reference (Figure 1). These idealized limits provide intuition for the behavior encountered in real prestack gathers and establish a baseline for interpreting phase variability in more complex settings. Figure 1 organizes the sequence from fully coherent to fully randomized phase behavior. In the noise-free limit (SNR → ∞), phases are aligned across traces, and the phase distribution is maximally concentrated. As noise is introduced, phase alignment degrades and the distribution broadens. In the opposite limit (SNR → −∞), the phases are uniformly distributed on [−π, π], indicating complete loss of coherence. The following subsections describe these three regimes in detail and use them to motivate circular statistical measures of phase direction and dispersion.

**Signal only (SNR = ∞)**

We begin with the noise-free reference case. Here the time-domain traces are perfectly repeatable within the local ensemble (Figure 1a), and the frequency-domain phases are coherently aligned across traces at all frequencies (Figure 1b). To describe this behavior statistically, we examine the distribution of phase values across the ensemble at each frequency, rather than individual phase spectra. For a multichannel dataset with N traces, define the set of phases at frequency ω as

$$\Theta(\omega) = \{\theta_1(\omega), \theta_2(\omega), \dots, \theta_N(\omega)\}. \tag{1}$$

In the absence of noise, all phases are identical, $\theta_1(\omega) = \theta_2(\omega) = \theta_N(\omega) = \theta_0(\omega)$, so the phase distribution collapses to a single value, which defines the lower bound of phase variability. This collapse is illustrated at 17 Hz and 52 Hz in Figures 1d and 1f, respectively, where phases are tightly concentrated, and the circular variance is zero (V = 0.00). This limiting case provides a baseline against which phase perturbations from additive noise and scattering-induced multiplicative distortions can be quantified.

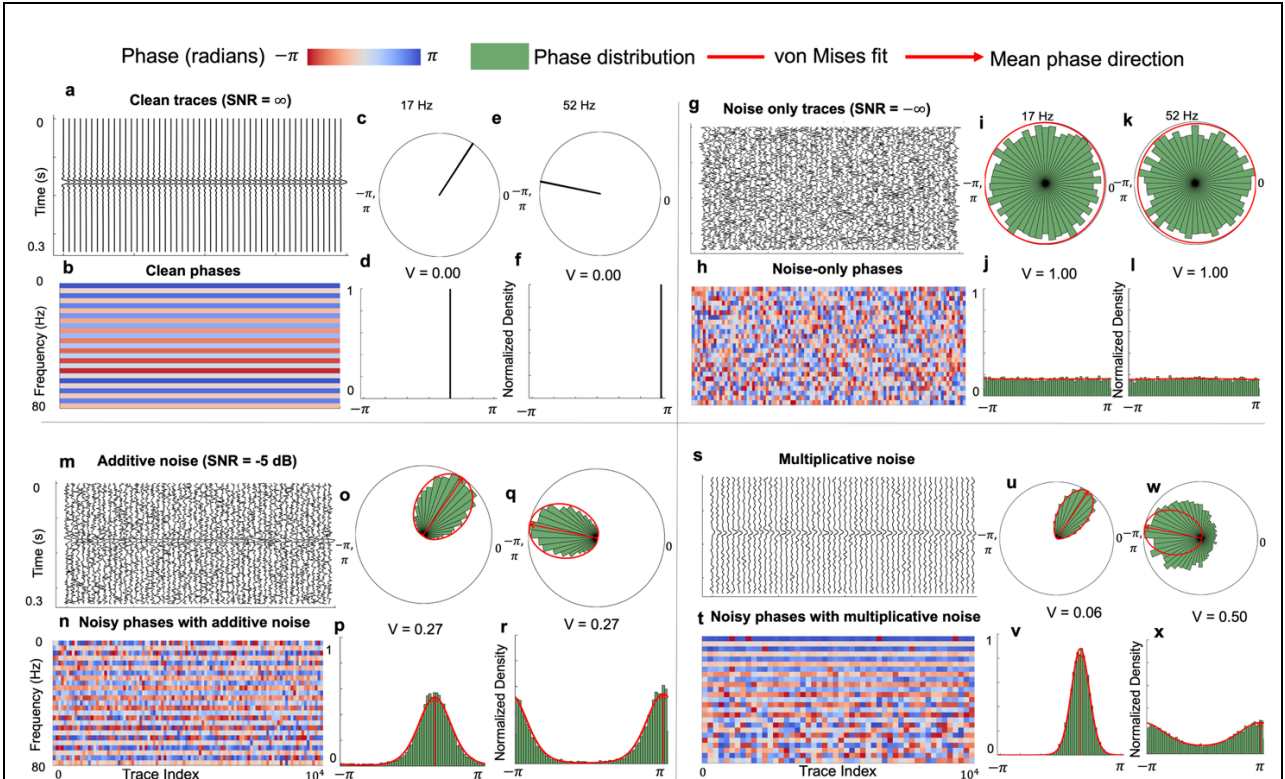

*Figure 1. Illustration of how phase behavior within local trace ensembles evolves from perfectly coherent to fully randomized as noise increases, and contrasts additive and multiplicative phase perturbations. Pure signal (SNR = ∞, panels a–f): (a) time-domain traces are repeatable, (b) phases are aligned across traces, (c, e) circular phase plots at 17 Hz and 52 Hz collapse to a single direction (d, f), and the corresponding linear histograms show zero dispersion, giving circular variance V = 0.00. Noise only (SNR = −∞, panels g–l): (g) time-domain traces are random, (h) phases are uncorrelated, (i, k) circular phase plots are uniform at both frequencies, and (j, l) linear histograms are flat, giving V = 1.00. Additive noise (SNR = −5 dB, panels m–r): (o, q) partial coherence remains but phase distributions broaden, (p, r) linear histograms at 52 Hz appear bimodal due to phase wrapping, with moderate dispersion (V ≈ 0.27). Multiplicative noise representative of scattering (panels s–x): (u, V = 0.06) phase variability becomes frequency dependent, with relatively stable low-frequency phases and (w, V = 0.50) strongly dispersed high-frequency phases, consistent with a transition toward speckle-like behavior.*

**Noise only (SNR = -∞)**

At the opposite extreme, we consider the noise-only limit, which defines the upper bound of phase variability (Figure 1g). With no coherent signal present, the frequency-domain phases are random and show no alignment across traces (Figure 1h). For a local ensemble of N traces, the phase set at frequency ω is again given by equation (1), but the phases are now mutually uncorrelated across channels. As illustrated in Figures 1i and 1k, the phases form a uniform circular distribution over [−π, π] at all frequencies,

$$\theta_k \sim U(-\pi, \pi), \qquad k = 1,2,\dots,N. \tag{2}$$

This case corresponds to a complete loss of phase coherence, with circular variance equal to one. Goodman (2007) showed that when a wavefield is modeled as a superposition of many independent scatterers with random phases, the resulting phase of the complex field is uniformly distributed on the circle. In the seismic context, Bakulin et al. (2022a) observed the same limiting behavior within a random multiplicative (speckle) framework, confirming that the noise-only end member exhibits uniform phase statistics.

**Noisy signal (-∞ < SNR < ∞)**

Most seismic data lie between the two end members and therefore contain both coherent reflections and noise. In this intermediate regime, reflection events are contaminated by either additive noise (Figure 1m) or multiplicative distortions associated with scattering (Figure 1s), so phase coherence is partially preserved but systematically perturbed, leading to broadened phase distributions within local ensembles.

As a conceptual model, we write the phase of the $k_{th}$ trace at frequency ω as

$$\theta_k(\omega) = \theta_0(\omega) + \delta_k(\omega), \qquad k = 1,2,\dots,N, \tag{3}$$

where $\theta_o(\omega)$ denotes the unperturbed signal phase and $\delta_k(\omega)$ is a symmetric random perturbation. As noise increases, the dispersion of $\{\theta_k(\omega)\}$ around $\theta_o(\omega)$ grows and the distribution broadens. To reproduce phase behavior observed in challenging land data, we adopt the multiplicative noise model of Bakulin et al. (2022a, 2023), which captures the key feature that scattering-induced phase perturbations increase with frequency; accordingly, phase distributions remain compact at low frequencies but broaden substantially at higher frequencies (Figures 1v and 1x). A practical complication is that seismic phase is wrapped to [−π, π], which makes linear summaries and linear histograms misleading, for example, through apparent bimodality caused purely by wrapping (Figures 1r and 1x). Although phase unwrapping is common in radar, InSAR, and optical interferometry (Zebker, 1998), unwrapping becomes unstable in noisy or incoherent regions (Feigl et al., 2007) and is problematic for seismic traces (Shatilo, 1992). These issues motivate treating seismic phase as circular data and analyzing it with circular statistics (Mardia and Jupp, 2000; Mohammad et al., 2021; Rohatgi et al., 2025). Circular statistics operate directly on wrapped phases, avoid boundary artifacts at −π and π, and provide robust measures of mean phase direction and dispersion, forming the basis for the phase-variance attributes introduced next (Bakulin et al., 2025; Rohatgi et al., 2025).

**CIRCULAR STATISTICS FOR SEISMIC PHASE ANALYSIS**

**Circular Mean : Measure of mean phase direction**

For an ensemble of seismic traces, the complex spectra can be normalized to unit magnitude so that only phase information is retained, independent of amplitude. This normalization ensures that phase statistics are not biased by variations in signal strength across traces. Figure 2 illustrates this procedure: noisy complex spectra (Figure 2a) are represented in the complex plane and then projected onto the unit circle (Figure 2b), preserving phase while removing amplitude effects.

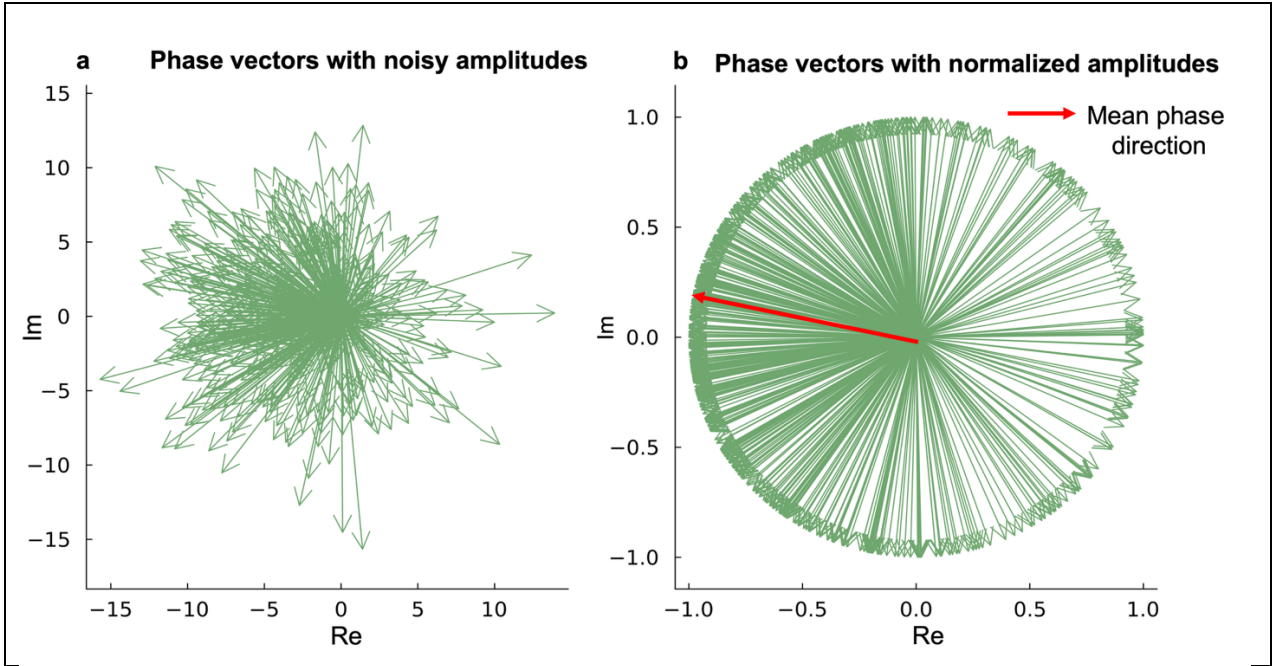

Figure 2. Phase vector representation at 52 Hz for a local ensemble of 500 traces with additive noise (SNR = −10 dB). Panel (a) shows complex phase vectors with their original amplitudes, where amplitude variability obscures the dominant phase direction. Panel (b) shows the same vectors after normalization to the unit circle, removing amplitude effects and revealing a well-defined mean phase direction that can be robustly characterized using circular statistics.

At a given frequency ω, the phase of each trace is represented as a unit phasor

$$e^{i\theta_1(\omega)}, e^{i\theta_2(\omega)}, \ldots e^{i\theta_N(\omega)}, \qquad (4)$$

Where for a complex signal $X_k(\omega)$, $\theta_k(\omega) = arg[X_k(\omega)]$ is the phase of the $k_{th}$ trace and N is the number of traces in the local ensemble.

The preferred phase direction is characterized by the circular mean $\bar{\theta}(\omega)$, defined as the direction of the resultant vector formed by summing the unit phasors. At a fixed frequency, the cosine and sine components are averaged as:

$$\bar{C}(\omega) = \frac{1}{N} \sum_{i=1}^{N} cos\theta_i(\omega), \ \bar{S}(\omega) = \frac{1}{N} \sum_{i=1}^{N} sin\theta_i(\omega). \qquad (5)$$

The length of the mean resultant vector is

$$\bar{R}(\omega) = \left(\bar{C}^2(\omega) + \bar{S}^2(\omega)\right)^{\frac{1}{2}}, \tag{6}$$

and the circular mean direction is given by

$$\bar{\theta}(\omega) = atan2\left(\bar{S}(\omega), \bar{C}(\omega)\right). \tag{7}$$

The mean direction is undefined when $\bar{R}(\omega) = 0$, corresponding to a uniform phase distribution with no preferred direction.

It is important to emphasize that the circular mean $\bar{\theta}$ is not equivalent to the arithmetic average $(\theta_1 + \theta_2 + \cdots + \theta_N)/N$. Instead, it represents the direction of the geometric vector sum of the unit phasors and correctly accounts for the periodic nature of phase. This distinction is critical when analyzing wrapped seismic phases, particularly in the presence of noise.

**Circular Variance: Measure of phase variability**

While the mean direction characterizes the central tendency of seismic phases within an ensemble, it is equally important to quantify how tightly the phases cluster around that direction. This variability is measured by the mean resultant length $\bar{R}$ (equation 6), which takes values between 0 and 1. Values of $\bar{R}$ close to 1 indicate strong phase coherence, whereas values near 0 indicate highly dispersed, incoherent phases,

$$0 \leq \bar{R}(\omega) \leq 1. \tag{8}$$

Although the mean resultant length is the fundamental quantity in circular statistics, it is often more intuitive, particularly by analogy with linear statistics, to express phase variability in terms of circular variance, defined as

$$V(\omega) = 1 - \bar{R}(\omega) \tag{9}$$

where, $0 \leq V(\omega) \leq 1$.

In the ideal signal-only case, all phase vectors are aligned and point in the same direction, yielding $\bar{R}(\omega) = 1$ and $V(\omega) = 0$, corresponding to perfect phase coherence (Figure 3a). In contrast, when the data consist solely of noise, phase vectors are uniformly distributed around the unit circle, resulting in $\bar{R}(\omega) \approx 0$ and $V(\omega) \approx 1$, indicating maximal phase variability (Figure 3b). Most realistic seismic situations lie between these two extremes. For example, with moderate additive noise at −5 dB, the phase vectors retain a preferred direction but exhibit noticeable dispersion, producing $\bar{R}(\omega) \approx 0.73$ and $V(\omega) \approx 0.27$ (Figure 3c). Under stronger or frequency-dependent phase perturbations, such as those associated with multiplicative noise, the phase vectors spread further and the circular variance increases accordingly, reaching values around $V(\omega) \approx 0.5$ (Figure 3d).

Although circular statistics provide a general framework for analyzing wrapped seismic phases, no specific parametric distribution is required to compute the mean direction or circular variance.

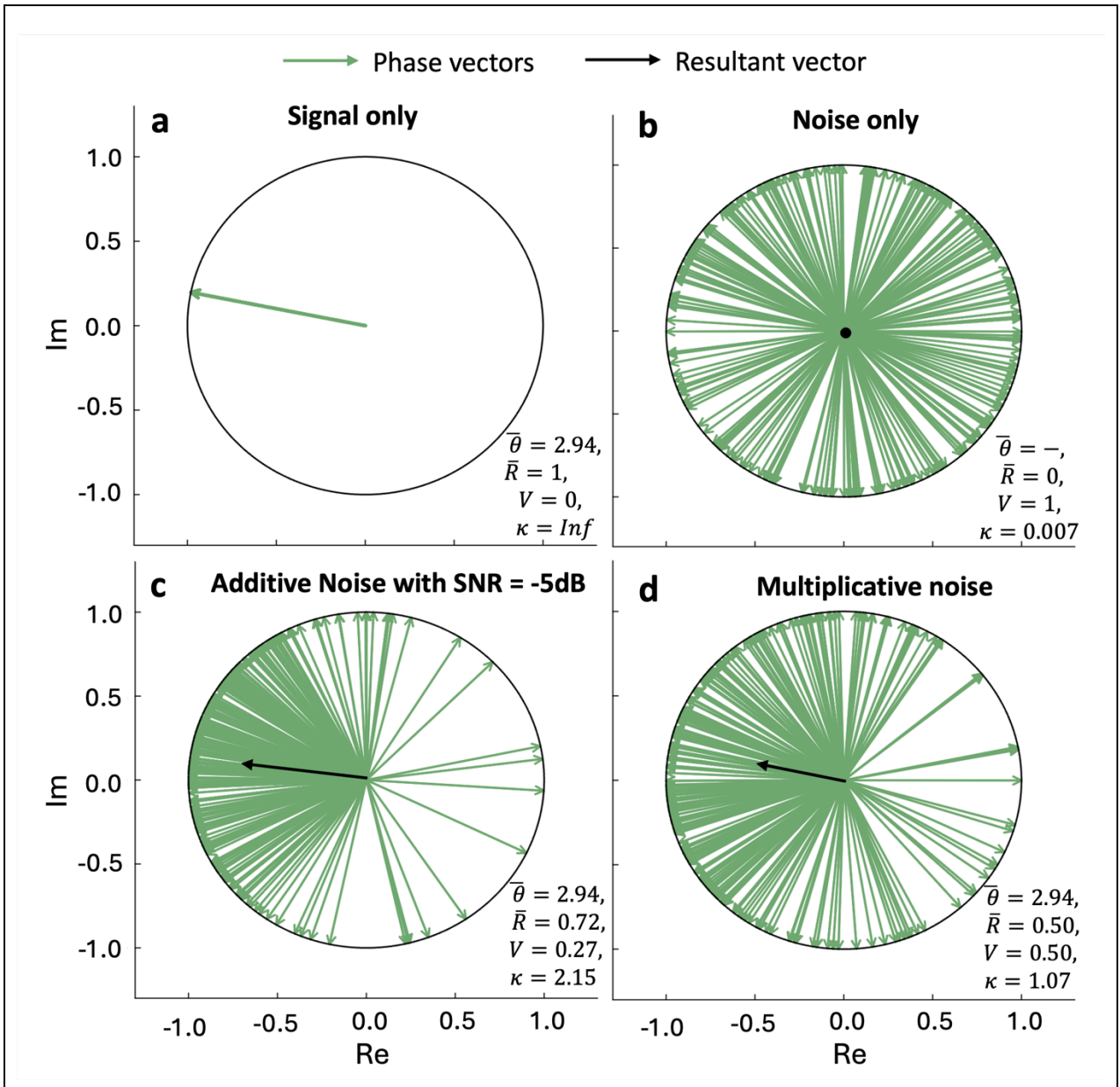

*Figure 3. Normalized phase vector representation at 52 Hz for an ensemble of 10,000 traces illustrating increasing phase variability. (a) Ideal signal case with no noise, where all phase vectors are aligned, giving a mean resultant length $\bar{R} = 1$ and zero circular variance $V = 0$. (b) Pure noise case, where phase vectors are uniformly distributed around the circle, yielding $\bar{R} = 0$ and maximum variance $V = 1$. (c) Intermediate case with additive noise at −5 dB, where phases retain a preferred direction defined by the circular mean but exhibit moderate dispersion, quantified by $V = 0.27$, with the black arrow indicating the mean resultant length $(1 - V)$. (d) Case with additional frequency-dependent phase perturbations representative of multiplicative noise, showing stronger dispersion and increased variance $V = 0.50$.*

In this study, the von Mises distribution is introduced as a convenient conceptual model, as it is the circular analogue of the normal distribution, providing an intuitive interpretation of phase clustering, dispersion, and a well-defined center of gravity when a single dominant signal event is present within the analysis window.

However, our analysis does not depend on the von Mises assumption. Instead, we primarily use the circular variance $V(\omega)$, which is defined directly from the mean resultant length and can be estimated nonparametrically from phase ensembles. Circular variance is a robust, model-free observable that remains stable under strong noise contamination, avoids phase-unwrapping ambiguities, and can be computed consistently across frequencies, offsets, and processing stages, making it well suited as a practical QC and diagnostic metric for seismic phase behavior.

**The von Mises Distribution**

The von Mises distribution, introduced by von Mises (1918) as the circular analogue of the normal distribution, has been widely used to model directional or wrapped data. It has found broad application in acoustics, optical interferometry, and InSAR, where phase is inherently periodic and must be analyzed without unwrapping (Feigl et al., 2009; Jiang et al., 2019; Takamichi et al., 2018; Nugraha et al., 2019). In these fields, von Mises models are commonly used to represent phase variability, reconstruct phase information, and analyze noisy or wrapped phase measurements in a probabilistically consistent manner. In seismology, Gosselin (2022) applied von Mises distributions to model uncertainty in circular phase spectra of surface-wave dispersion, while Rohatgi et al. (2025) used circular statistics to quantify phase coherence within local seismic ensembles.

This distribution is parameterized by the mean direction $\bar{\theta}$ and the concentration parameter $\kappa$. Its probability density function given as:

$$f(\theta; \bar{\theta}, \kappa) = \frac{1}{2\pi I_0(\kappa)} e^{\kappa \cos(\theta - \bar{\theta})}, \qquad (10)$$

where $I_0$ denotes the modified Bessel function of the first kind and order zero, defined as

$$I_0(\kappa) = \frac{1}{2\pi} \int_0^{2\pi} e^{\kappa \cos\theta} d\theta. \qquad (11)$$

The function $I_0$ also admits the power series expansion

$$I_0(\kappa) = \sum_{r=0}^{\infty} \frac{1}{(r!)^2} \left(\frac{\kappa}{2}\right)^{2r}. \qquad (12)$$

The von Mises distribution is unimodal and symmetric about its mean direction. When $\kappa = 0$, it reduces to the uniform distribution on [–π, π], corresponding to completely random phases. As $\kappa$ increases, the distribution becomes increasingly concentrated around $\bar{\theta}$ and approaches a normal distribution in shape, while preserving phase periodicity. Figure 4a illustrates this behavior for a fixed mean direction of 2.2 radians and increasing $\kappa$, showing the transition from nearly uniform to strongly clustered phase distributions.

The concentration parameter $\kappa$ can be estimated from the mean resultant length $\bar{R}$ using standard approximations (Fisher, 1993),

$$\kappa = \begin{cases} 2\bar{R} + R^3 + \dfrac{5R^5}{6} &, \quad \bar{R} < 0.53 \\ -0.4 + 1.39\bar{R} + \dfrac{0.43}{1-\bar{R}} &, \quad 0.53 \leq \bar{R} < 0.85 \\ \dfrac{1}{3R - 4R^2 + R^3} &, \quad \bar{R} \geq 0.85 \end{cases} \qquad (13)$$

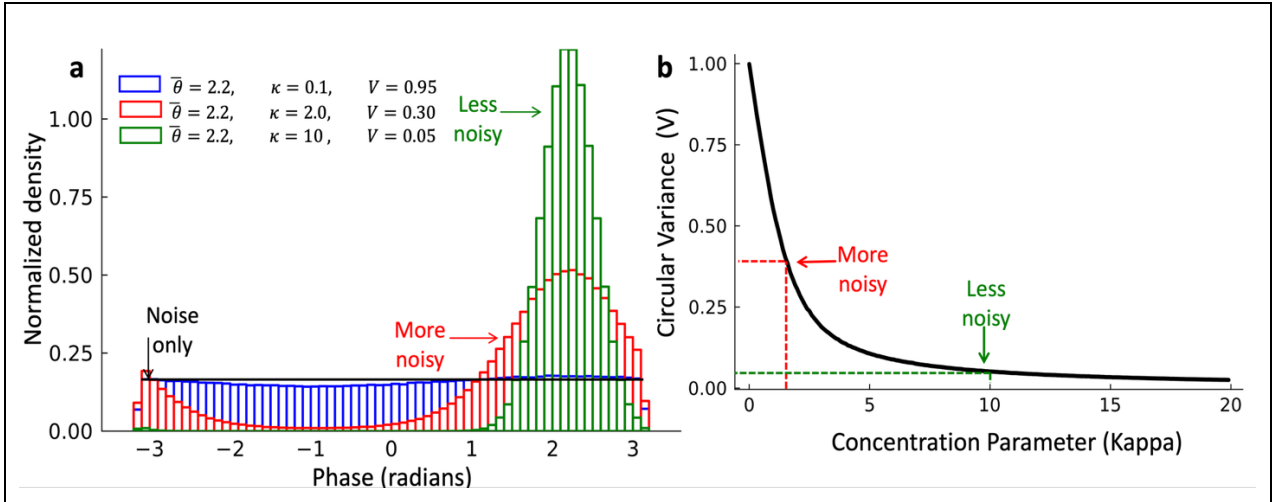

Figure 4. (a) Histograms of synthetic phases sampled from von Mises distributions with fixed circular mean $\bar{\theta} = 2.2$. For low concentration ($\kappa = 0.1$), the distribution is nearly uniform with high variance. As $\kappa$ increases to 2 (red), the variance decreases and phases begin to cluster around the mean. For higher concentration ($\kappa = 10$), phases are strongly concentrated along the mean, and the variance reduces further to 0.05. (b) Relationship between circular variance and concentration parameter $\kappa$. This plot illustrates the nonlinear mapping between circular variance (V) and the von Mises concentration parameter ($\kappa$), derived numerically via the mean resultant length $\bar{R}$. As $\kappa$ increases, V decreases rapidly, indicating tighter phase clustering. The curve highlights that at low variance (high coherence), small changes in V correspond to large swings in $\kappa$, underscoring why V is often more stable and interpretable in noisy seismic data.

Circular variance $V$ and the von Mises concentration parameter $\kappa$ both describe the dispersion of phase vectors around the mean direction, but from complementary perspectives. Circular variance is defined directly from the mean resultant length $V = 1 - \bar{R}$ and provides a bounded, geometrically intuitive measure of phase variability, where values near zero indicate strong phase coherence and values near one indicate incoherent or noise-dominated behavior. In contrast, $\kappa$ is specific to the von Mises model and quantifies the strength of clustering in an unbounded manner, ranging from zero to infinity. As a result, $\kappa$ can be more sensitive to noise and may fluctuate strongly between neighboring ensembles, particularly at low coherence. When the von Mises model is assumed, however, $V$ and $\kappa$ are uniquely related through $\bar{R}$ and can be converted analytically or numerically, as illustrated in Figure 4b. For clarity, Table 1 summarizes the key circular statistical quantities used in this study and their physical interpretation.

| Attribute | Symbol | Range | Interpretation |
|---|---|---|---|
| Circular mean | $\bar{\theta}$ | $[-\pi, \pi]$ | Mean (central) phase direction |
| Mean resultant length | $\bar{R}$ | $[0, 1]$ | Degree of phase alignment (clustering strength) |
| Circular variance | $V = 1 - \bar{R}$ | $[0, 1]$ | Phase dispersion or incoherence |
| Concentration (von Mises) | $\kappa$ | $[0, \infty)$ | Phase clustering strength (inverse dispersion) |

Table 1: Summary of key circular statistical attributes used for ensemble-based seismic phase analysis.

**QUANTIFYING VARIABILITY IN SEISMIC PHASE**

Although phase statistics are estimated directly from the data, without fitting any specific parametric distribution, the von Mises distribution remains an important interpretive reference. When a single coherent reflection dominates a local ensemble, the phase distribution is expected to be unimodal, with the mean direction representing the signal phase and the dispersion reflecting noise. This provides a simple and intuitive signal–plus–noise picture in circular phase space.

Real seismic data are more complex. Noise can itself be partially coherent, such as ground roll, guided waves, or other organized wave modes, and multiple signals may interfere or cross within the same analysis window. In such cases, phase distributions may broaden asymmetrically or become multimodal. This behavior does not merely indicate stronger noise, but rather the absence of a single dominant signal phase. Importantly, the apparent phase distribution may depend on the number of traces in the ensemble: small local ensembles can appear unstable or noisy, whereas larger ensembles tend to reveal more stable phase structure. Even under these conditions, the unimodal von Mises case remains a useful reference state, providing a baseline for distinguishing signal-dominated behavior from mixed or incoherent phase regimes.

Building on this conceptual framework, we demonstrate how circular variance can be used as a quantitative quality-control attribute for seismic phase. Circular variance measures the spread of phase distributions within a local ensemble. Because seismic phase variability depends on both offset and frequency, these statistics must be computed locally using sliding windows over ensembles of traces. This approach enables objective tracking of phase variability across a seismic gather and provides a means to assess the impact of processing steps.

The workflow for computing circular variance is summarized as follows. Each local time-space window is analyzed in the frequency domain using the following procedure.

1. Each trace $x_k(t)$ within the window is transformed to the frequency domain,
$X_k(\omega) = \mathcal{F}\{x_k(t)\} = |X_k(\omega)| e^{i\theta_k(\omega)}$,
where $|X_k(\omega)|$ is the spectral amplitude and $\theta_k(\omega)$ the spectral phase.

2. To isolate phase information, each spectrum is normalized to unit magnitude,
   $\tilde{X}_k(\omega) = \frac{X_k(\omega)}{|X_k(\omega)|} = e^{i\theta_k(\omega)}.$
   The set $\{\tilde{X}_k(\omega)\}_{k=1}^N$ thus forms a phase ensemble at frequency $\omega$.

3. Phase dispersion within the local ensemble is quantified using the circular variance,
   $V(\omega) = 1 - |\frac{1}{N}\sum_{k=1}^N e^{i\theta_k(\omega)}|,$
   where $V(\omega) \in [0,1]$.

4. The phase variance $V(\omega)$ is assigned to the center of the analysis window and mapped across time and frequency.

The ensemble size N is chosen to ensure statistically stable and interpretable estimates, as discussed in Appendix A. To assess the reliability of this metric, we first apply it to a controlled synthetic example. This allows us to demonstrate how circular variance responds to known phase perturbations, providing a benchmark before extending the method to more complex field data.

**Synthetic example with phase perturbations**

The choice of ensemble size is dictated by the role of the analysis and is kept consistent between synthetic tests and field applications. In high-noise conditions, larger ensembles are required to see through noise and obtain statistically stable estimates, whereas in low-noise settings smaller ensembles may suffice. Accordingly, large ensembles of 10,000 traces analyzed using 2,000-trace sliding windows are employed to obtain statistically stable estimates of circular variance when it is used as a diagnostic and QC attribute. This ensures that the measured phase variability reflects the imposed perturbations rather than sampling noise.

To investigate the variability of seismic phase within an ensemble, we designed a controlled synthetic experiment. Phase perturbations were introduced by drawing random realizations from a von Mises distribution, consistent with the conceptual model described earlier. For each trace, a circular variance V was prescribed and systematically varied across the ensemble to impose laterally varying phase noise. Figure 5a illustrates the imposed variation in phase perturbations as a function of trace number. The variance decreases from left to right, mimicking a realistic acquisition scenario in which near-offset traces fall within a noise cone characterized by stronger phase perturbations, while far-offset traces exhibit increasingly coherent reflection behavior. The resulting synthetic seismic section is shown in Figure 5b.

The baseline signal was generated using a Klauder wavelet, defined as the autocorrelation of a vibroseis sweep commonly employed in land seismic surveys. This wavelet served as a clean reference, from which an ensemble of 10,000 identical traces was constructed. In the frequency domain, the unperturbed signal at channel $k$ and frequency $\omega$ is given by

$$X_k(\omega) = A_0(\omega)e^{i\theta_0(\omega)}, \tag{14}$$

where $A_0(\omega)$ denotes the true amplitude spectrum and $\theta_0(\omega)$ the corresponding true phase. Phase distortions were then introduced by applying spatially dependent perturbations, such that the perturbed signal becomes

$$X_k(\omega) = A_0(\omega)e^{i\{\theta_0(\omega) + \theta_k{'}(\omega)\}}, \tag{15}$$

where $\theta_k'(\omega)$ represents the imposed perturbation.

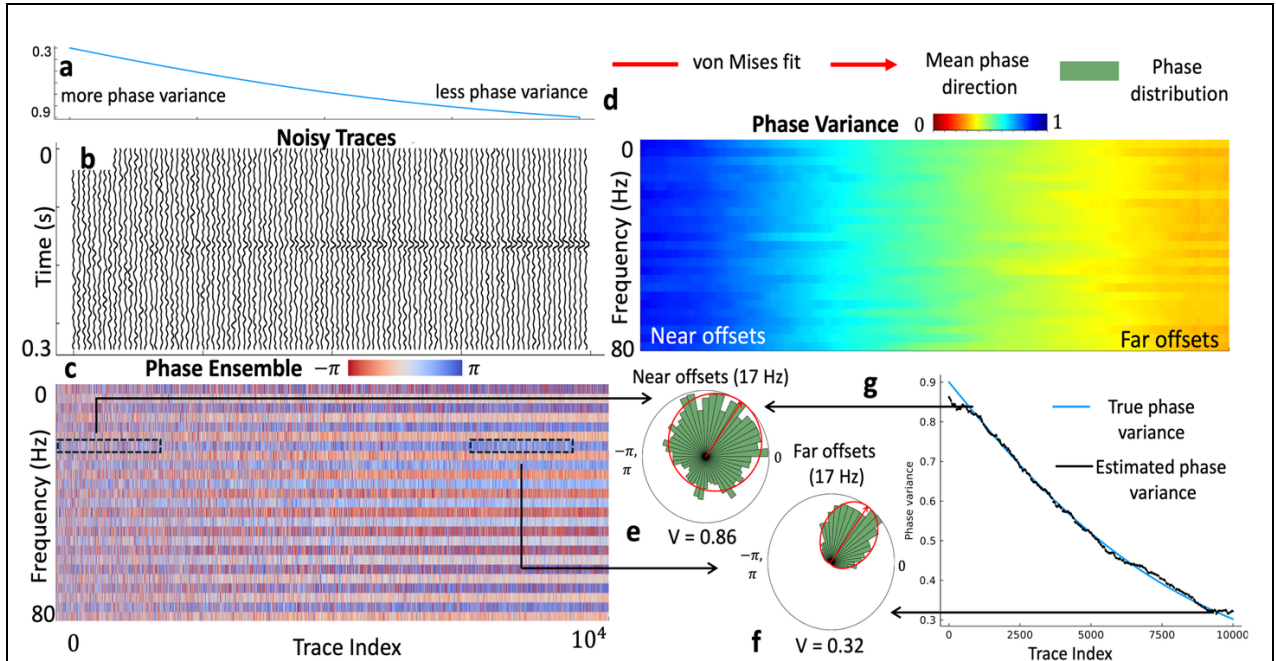

*Figure 5. Quantifying phase perturbations in a controlled synthetic example. Panel (a) shows the prescribed circular variance varying smoothly with trace index, representing decreasing phase perturbations from near to far offsets. Panel (b) displays the corresponding ensemble of 10,000 synthetic traces, with strong phase distortion at near offsets and increasingly coherent reflections at far offsets. Panel (c) shows the phase ensemble in the frequency domain after amplitude normalization. Panel (d) presents the phase-variance map computed using 2,000-trace sliding windows, illustrating the spatial and spectral evolution of phase variability across the gather. Panels (e) and (f) show phase distributions from representative near-offset and far-offset windows at 17 Hz, with circular variance values of 0.86 and 0.32, respectively, highlighting the contrast between noise-dominated and more coherent phase behavior. Panel (g) compares the imposed and estimated phase variance as a function of trace index, demonstrating close agreement and validating the reliability of circular variance as a measure of seismic phase perturbations.*

Within each window, amplitudes were normalized to isolate phase effects, ensuring that the computed statistics reflect phase variability alone. The resulting phase ensemble in the frequency domain is shown in Figure 5c. Sliding the window across the ensemble allows phase distributions to be evaluated as a function of offset and frequency. Representative phase distributions extracted from near-offset and far-offset windows are shown in Figures 5e and 5f, respectively. These distributions illustrate the transition from highly dispersed phases at near offsets to more concentrated phase behavior at far offsets. For example, at 17 Hz the near-offset window yields a circular variance of 0.86, indicating strong phase variability, whereas the far-offset window produces a lower value of 0.32, reflecting greater phase stability.

The circular variance computed within each window is assigned to the window center before advancing to the next position. Repeating this procedure produces a new seismic attribute, referred to here as phase variance. The resulting phase-variance map (Figure 5d) captures the spatial and spectral evolution of phase variability across the synthetic gather. As expected, phase variance is highest at near offsets and decreases steadily toward far offsets, consistent with the imposed perturbations. Because the perturbations in this test were identical across frequencies, the frequency dependence of the phase-variance map is uniform. The framework naturally extends to cases where phase noise varies with frequency, although the present example focuses on spatial variability for clarity.

The average phase variance across all traces (Figure 5g) shows close agreement between the imposed and recovered values, confirming that circular variance reliably quantifies phase perturbations in a controlled setting. These synthetic results demonstrate that circular variance provides a robust and interpretable measure of seismic phase variability. Having validated that $V(\omega)$ accurately recovers known phase perturbations in synthetic data, we now apply it as a QC observable to field seismic gathers to diagnose where phase becomes unreliable.

**Monitoring phase variance in field data**

Monitoring phase quality is critical in seismic processing because phase integrity directly controls imaging, inversion, and attribute analysis. Despite its importance, conventional workflows provide few tools for quantifying phase variability or tracking phase noise in a systematic way. As a result, phase distortions caused by near-surface effects, scattering, or processing artifacts may persist and propagate through the workflow without being explicitly diagnosed. Circular variance provides a practical phase quality-control attribute that allows phase behavior to be evaluated objectively as a function of offset and frequency.

To illustrate this capability, we analyze two prestack seismic gathers extracted from a 3D CMP supergather, one before and one after conventional time processing. A 300-ms time window containing a deep reflection event was selected. Phase statistics were computed using a sliding-window approach with local ensembles of 2,000 traces across a total of 10,000 traces. The gathers shown in Figures 6a and 6b are displayed after normal moveout correction and statics. The conventional time-processing sequence included linear noise attenuation, refraction statics, random noise removal, two passes of surface-consistent deconvolution, post-deconvolution noise attenuation targeting linear, random, and burst noise, merge-phase and static matching, surface-consistent scaling, velocity analysis, and residual statics correction.

The effect of processing on seismic phase can be examined by inspecting phase distributions within local time–frequency windows. At 16 Hz in a mid-offset local window, the pre-processed data exhibit a broad and weakly organized phase distribution, indicative of substantial phase variability (Figure 6c). After conventional processing, the same window shows noticeably tighter phase clustering (Figure 6d), with circular variance decreasing from 0.81 to 0.59. This reduction reflects improved phase coherence at that frequency and offset and demonstrates how phase variance provides a quantitative measure of processing impact on seismic phase behavior in field data.

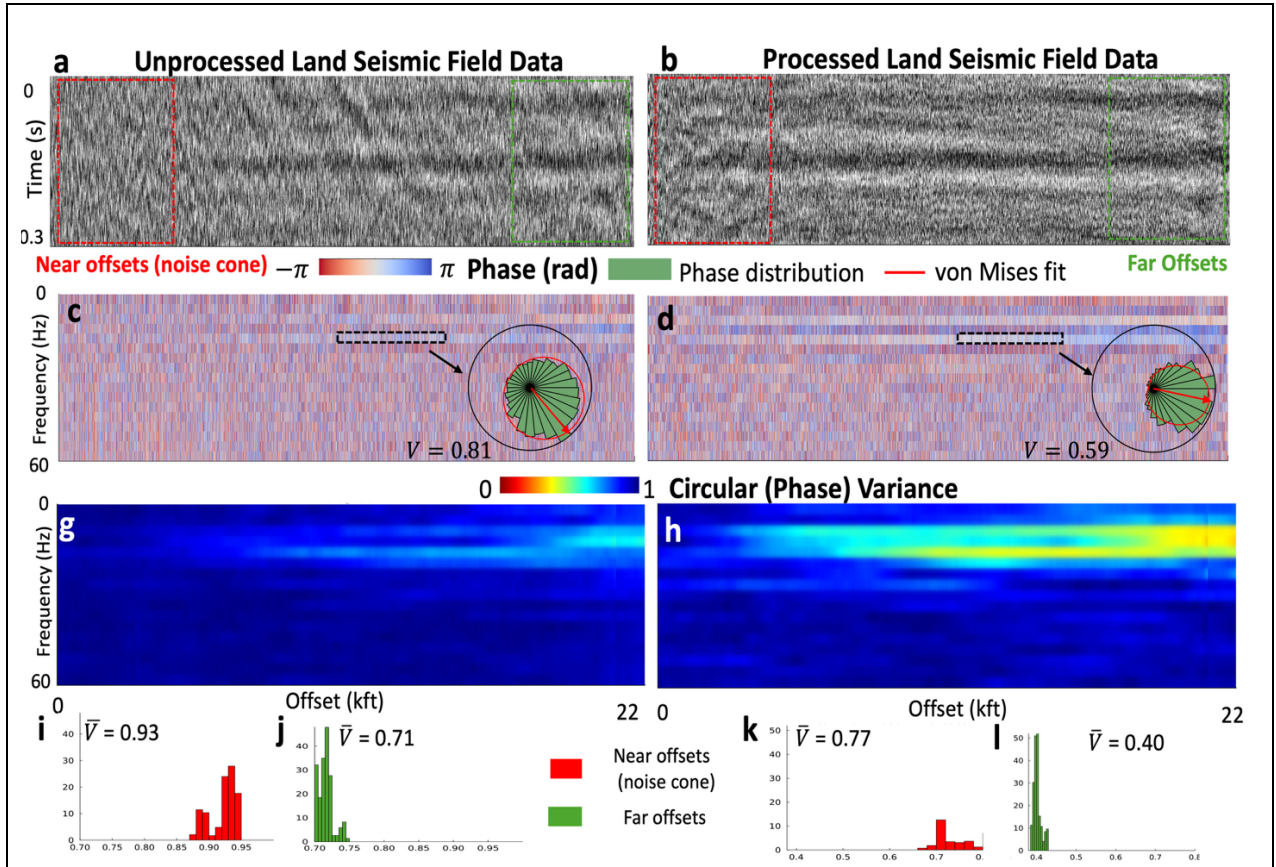

Figure 6. Phase-based quality control on prestack land data before and after conventional time processing. (a, b) Seismic gathers after NMO and statics, with near offsets inside the noise cone and far offsets marked for reference. Visual continuity improves after processing, but this alone does not quantify phase reliability. (c, d) Local phase distributions at mid offsets and 16 Hz show strong phase disorder before processing (V = 0.81) and tighter clustering after processing (V = 0.59), indicating improved but still imperfect phase coherence. (e, f) Corresponding phase ensembles illustrate how processing narrows the phase spread rather than simply boosting amplitudes. (g, h) Frequency–offset maps of circular variance reveal where processing is effective: phase variability is reduced mainly at mid to far offsets and lower frequencies, while near offsets and higher frequencies remain noise dominated. (i–l) Histograms of circular variance at 16 Hz quantify this behavior, showing that processing lowers V from noise-like values near 0.9 toward more coherent values around 0.4–0.6. Together, these panels show that phase variance provides a direct, operational measure of where seismic phase becomes reliable after processing, beyond what can be inferred from visual inspection or amplitude alone.

To generalize these observations, we computed circular variance across all frequencies and offsets using overlapping sliding time windows (Figures 6g and 6h). Although reflectors appear visually clearer after conventional processing, the phase-variance maps reveal a more nuanced picture. At higher frequencies, circular variance remains close to one, indicating that phase behavior is still dominated by noise despite apparent amplitude enhancement. The spatial distribution of variance highlights systematic trends. Near offsets, which fall within the noise cone, exhibit consistently high phase variance, while mid to far offsets show progressively lower variance. For example, at 16 Hz, the unprocessed near-offset data yield a variance of 0.93, characteristic of noise-dominated phase behavior, whereas mid to far offsets decrease to about 0.71, indicating partial coherence. After processing, these same offsets show a marked reduction in variance to approximately 0.40, quantitatively confirming that conventional processing improves phase coherence primarily at lower frequencies and away from the noise cone (Figures 6j and 6l).

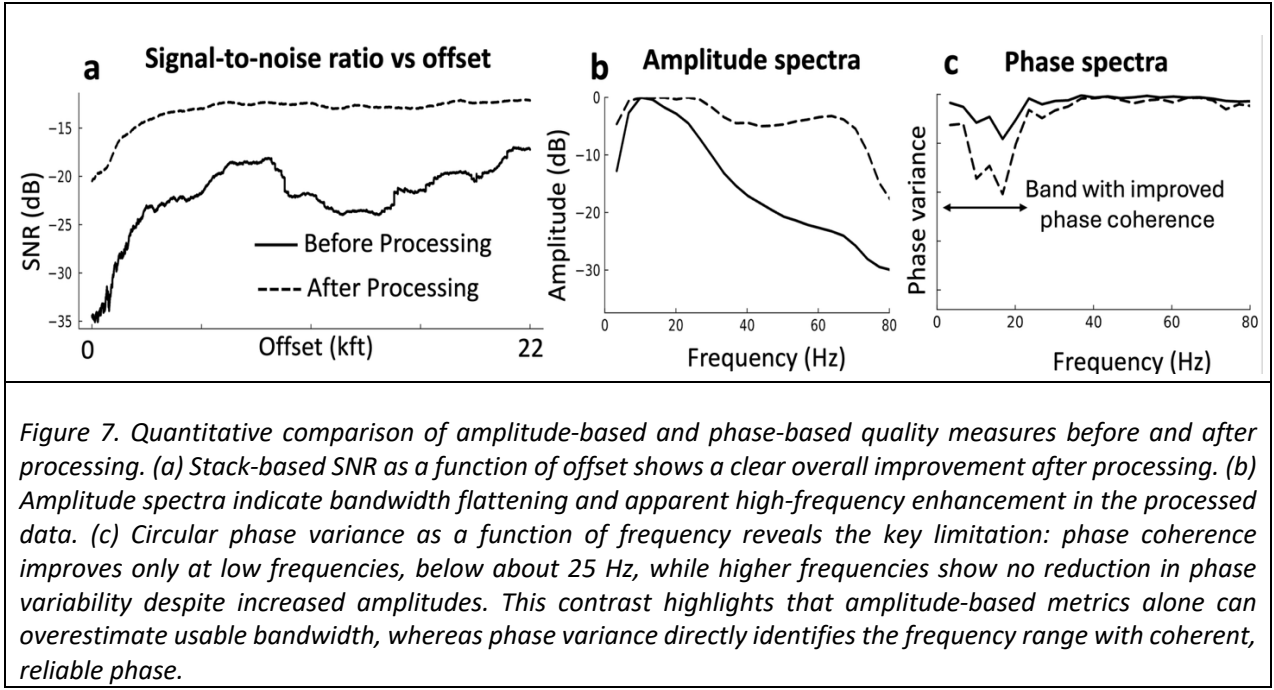

*Figure 7. Quantitative comparison of amplitude-based and phase-based quality measures before and after processing. (a) Stack-based SNR as a function of offset shows a clear overall improvement after processing. (b) Amplitude spectra indicate bandwidth flattening and apparent high-frequency enhancement in the processed data. (c) Circular phase variance as a function of frequency reveals the key limitation: phase coherence improves only at low frequencies, below about 25 Hz, while higher frequencies show no reduction in phase variability despite increased amplitudes. This contrast highlights that amplitude-based metrics alone can overestimate usable bandwidth, whereas phase variance directly identifies the frequency range with coherent, reliable phase.*

This trend is consistent with stack-based signal-to-noise ratio estimates (Figure 7a), computed following Bakulin et al. (2022b). Prestack SNR improves from roughly −22 dB to −12 dB after processing, reflecting substantial suppression of random noise. However, SNR averages energy over the full frequency band and therefore masks important broadband behavior. In contrast, circular variance isolates phase variability directly and reveals that improvements are confined to a limited frequency range.

Amplitude spectra before and after processing (Figure 7b) show that conventional workflows often boost high frequencies through repeated deconvolution and filtering, giving the appearance of increased bandwidth. Amplitude-based measures alone, however, do not capture phase stability and can therefore be misleading. Phase-variance spectra (Figure 7c) provide a complementary and more physically meaningful perspective. By examining circular variance as a function of frequency, averaged over offsets, it becomes possible to define an effective frequency band in terms of phase coherence rather than amplitude alone. In practice, one can select a threshold variance corresponding to acceptable phase stability for a given application. Frequencies below this threshold retain coherent phase and are suitable for phase-sensitive workflows such as full-waveform inversion and AVO analysis, whereas frequencies above it contributes primarily incoherent energy.

These observations suggest that seismic processing should be evaluated not only by amplitude and SNR metrics but also by its ability to systematically reduce phase variance. Monitoring phase variance provides a direct means to assess whether a given processing step improves or degrades phase integrity and offers a quantitative guide for parameter selection. Although we examine only two stages here, raw data and fully time-processed data, the same framework can be applied step by step throughout the processing sequence.

Identifying frequency bands with acceptable phase variance naturally leads to strategies for mitigating phase variability within those bands. In practice, seismic processing aims not only to diagnose phase distortions but also to compensate for them. Bakulin et al. (2023) showed that when the underlying signal is consistent across traces, the phase of the clean signal can be recovered using phase masking based on local ensembles, because the phase spectrum of the ensemble expectation matches that of the signal. In this case, the circular mean phase provides a convenient, amplitude-independent estimate of the signal phase.

**DISCUSSION**

Treating seismic phase as a circular variable reframes phase variability from a difficult, trace-level attribute into a measurable ensemble property that can be evaluated systematically across frequency and offset. Circular variance, here called phase variance $V(\omega)$, operates directly on wrapped phases and avoids the ambiguities of phase unwrapping. Low $V(\omega)$ corresponds to tightly clustered phases and coherent signal behavior, whereas high $V(\omega)$ indicates strong phase dispersion caused by noise, scattering, or other incoherent effects. In addition, the circular mean phase provides a useful estimate of the underlying signal phase within an ensemble. Although not the focus of this study, this estimate may support additional quality control of phase trends and serve as a key ingredient in phase-enhancement approaches such as seismic time-frequency masking (Bakulin et al., 2023) and phase-based filtering.

Synthetic tests confirm that $V(\omega)$ reliably recovers imposed phase variability as a function of frequency and offset. Application to field data shows that conventional time-domain processing primarily reduces phase variance at low frequencies and mid-to-far offsets. At higher frequencies, phase variance often remains elevated, even when amplitude spectra suggest apparent bandwidth extension. This discrepancy highlights a key limitation of amplitude-based diagnostics and motivates defining an effective bandwidth in terms of phase coherence rather than amplitude alone.

Phase variance maps and spectra provide a direct way to identify where phase information becomes reliable after processing. By selecting an application-dependent threshold on $V(\omega)$, practitioners can objectively delineate the frequency range suitable for phase-sensitive workflows, such as AVO analysis, full-waveform inversion, and migration. Unlike visual inspection or SNR metrics, $V(\omega)$ isolates phase behavior and reveals frequency-dependent limitations that would otherwise remain hidden. The statistical stability of $V(\omega)$ and its dependence on ensemble size are examined in Appendix A.

**CONCLUSIONS**

We have introduced circular statistics as a new and necessary framework for analyzing seismic phases. By treating phase as a circular variable rather than a trace-level artifact, phase variability becomes a measurable ensemble property that can be quantified, tracked, and acted upon systematically. Circular variance is not merely another quality-control attribute but a physically grounded descriptor of phase coherence that directly reflects the reliability of phase information across frequency and offset.

Our results show that amplitude-based diagnostics alone can substantially overestimate usable bandwidth. Conventional processing often amplifies high-frequency amplitudes while leaving phase behavior noise-dominated. Phase variance directly exposes this limitation, enabling the definition of an effective bandwidth based on phase coherence rather than spectral amplitude. This represents a fundamental shift in how seismic data quality is assessed for phase-sensitive applications.

We propose using phase variance as a standard prestack diagnostic, evaluated alongside amplitude spectra and SNR. Tracking $V(\omega)$ through a processing sequence provides an objective way to identify where phase integrity improves or deteriorates, to guide parameter selection, and to prevent the propagation of phase distortions that would otherwise remain invisible. Importantly, this diagnostic view naturally enables phase-conditioning and masking strategies driven by explicit objectives, such as achieving a prescribed level of phase fidelity for imaging or inversion, rather than by visual QC or amplitude-based criteria.

More broadly, this work establishes circular statistics as a practical and extensible foundation for seismic phase analysis and provides the statistical basis for future phase-enhancement and masking approaches that operate directly on ensemble phase behavior. The framework applies equally to QC, processing design, and phase-sensitive imaging and inversion. We anticipate that phase-based diagnostics and conditioning methods built on circular statistics will become increasingly important as seismic workflows move toward higher frequencies, denser sampling, and greater sensitivity to subtle phase effects.

# APPENDIX A: EFFECT OF ENSEMBLE SIZE ON THE STABILITY OF PHASE VARIANCE ESTIMATE

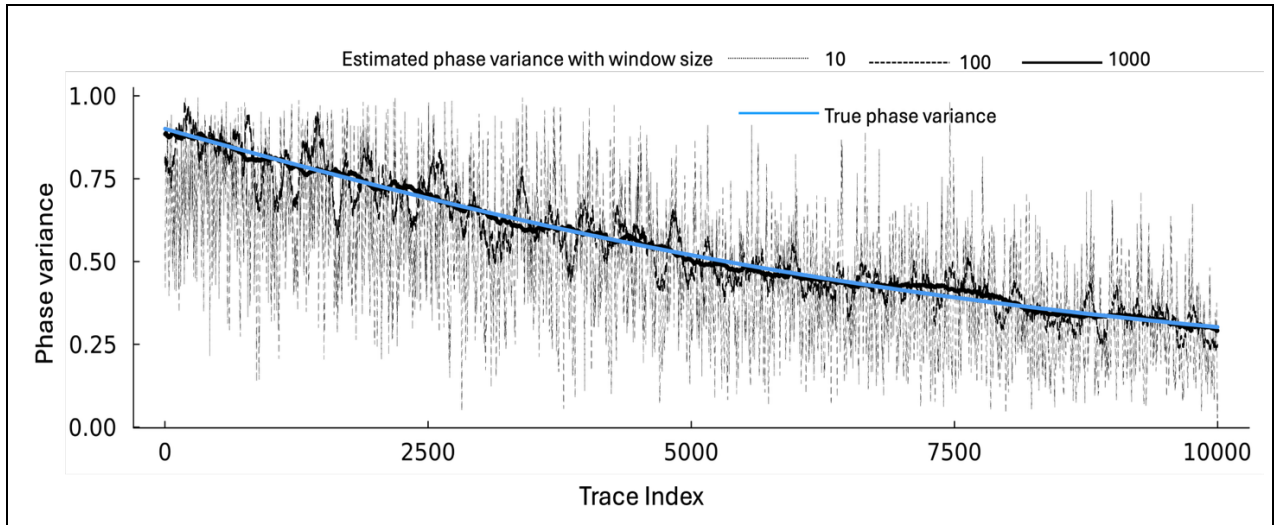

**Figure A1.** *Estimated circular phase variance computed using sliding windows of 10, 100, and 1000 traces. The blue curve shows the true imposed phase variance, while the black curves show estimates obtained with different ensemble sizes. All window sizes recover the same large-scale trend in phase variability, indicating that the underlying phase behavior is captured regardless of ensemble size. Differences appear only in the fine-scale fluctuations: smaller ensembles exhibit higher estimator variability, while larger ensembles yield smoother and more statistically stable variance estimates. This demonstrates that ensemble size controls the stability of the estimate, not the inferred phase trend itself.*

Circular variance is a statistical quantity estimated from a finite ensemble of phase samples. As such, the variance itself is a random variable whose stability depends on the number of traces included in the ensemble. Importantly, small ensemble sizes do not render phase-variance estimates uninterpretable; rather, they increase estimator variability and make interpretation more sensitive to local fluctuations. Interpretable phase-variance estimates therefore require an ensemble size appropriate to the objective of the analysis. To assess the effect of ensemble size, we computed circular variance using sliding windows containing 10, 100, and 1000 traces (Figure A1). For small ensembles, the estimated variance exhibits strong local fluctuations due to limited sampling, revealing fine-scale detail but with increased statistical scatter. . As the number of traces increases, these fluctuations are progressively suppressed, and the estimated variance converges toward a smooth and stable trend that reflects the underlying phase dispersion rather than estimator noise. Importantly, the large-scale trend in phase variability is preserved across all window sizes, while increasing the ensemble size improves statistical stability.

This behavior is expected and reflects the nested nature of the problem: individual phase values are random, and circular variance is itself an estimate derived from those random samples. Consequently, the ensemble size must be chosen based on the analysis objective. When the goal is to resolve only the underlying trend in phase variability, large ensembles are required, particularly when circular variance is used as a diagnostic or quality-control attribute where stability and robustness are critical. Smaller ensembles may be useful when emphasizing localized structure, but at the expense of increased variability and interpretational uncertainty.

      This distinction underpins the consistent use of different ensemble sizes throughout this study and ensures that circular variance is applied appropriately to each task.